\documentclass[prb,twocolumn,superscriptaddress,showpacs,floatfix]{revtex4}

\usepackage{color} 
\usepackage{tabularx} 
\usepackage{epsfig}
\usepackage{amsmath} 
\usepackage{amssymb} 
\usepackage{graphicx}
\usepackage{wasysym}
\usepackage{oldgerm}
\usepackage{psfrag}
\usepackage{dcolumn}
\usepackage{bm}

\begin{document}

\title{Diluted antiferromagnets in a field seem to be in a different
universality class than the random-field Ising model}

\author{Bj\"orn Ahrens}
\affiliation{Department of Physics and Astronomy, Texas A\&M University, 
College Station, Texas 77843-4242, USA}
\affiliation{Institut f\"ur Physik, Carl-von-Ossietzky Universit\"at, 
26111 Oldenburg, Germany}

\author{Jianping Xiao}
\affiliation{Department of Physics and Astronomy, Texas A\&M University,
College Station, Texas 77843-4242, USA}

\author{Alexander K. Hartmann}
\affiliation{Institut f\"ur Physik, Carl-von-Ossietzky Universit\"at, 
26111 Oldenburg, Germany}

\author{Helmut G. Katzgraber}
\affiliation{Department of Physics and Astronomy, Texas A\&M University,
College Station, Texas 77843-4242, USA}
\affiliation{Materials Science and Engineering Program, Texas A\&M
University, College Station, TX 77843-3003, USA}
\affiliation{Theoretische Physik, ETH Zurich, CH-8093 Zurich, Switzerland}

\date{\today}

\begin{abstract}

We perform large-scale Monte Carlo simulations using the
Machta-Newman-Chayes algorithms to study the critical behavior of both
the diluted antiferromagnet in a field with 30\% dilution and the
random-field Ising model with Gaussian random fields for different field
strengths. Analytical calculations by Cardy [Phys.~Rev.~B {\bf 29}, 505
(1984)] predict that both models map onto each other and share the same
universality class in the limit of vanishing fields. However, a detailed
finite-size scaling analysis of the Binder cumulant, the two-point
finite-size correlation length, and the susceptibility suggests that even
in the limit of small fields, where the mapping is expected to work,
both models are not in the same universality class.  Based on our
numerical data, we present analytical expressions for the phase
boundaries of both models.

\end{abstract}

\pacs{64.60.De, 75.10.Nr, 75.40.-s,75.50.Lk}

\maketitle

\section{\label{Introduction}Introduction}

The random-field Ising model\cite{belanger:91} (RFIM) is of paramount
importance in the field of disordered
systems.\cite{nattermann:88,belanger:98,nattermann:98,lyuksyutov:99} A
plethora of problems across disciplines can be studied via the RFIM,
ranging from the thermodynamics of disordered magnets,\cite{young:98}
hysteresis in magnetic systems and Barkhausen
noise,\cite{sethna:93,perkovic:95,perkovic:99} tunable domain-wall
pinning,\cite{silevitch:10} the random pinning of
polymers,\cite{halpinhealy:91} and even water seepage in porous media.
As such, the RFIM is still under intense theoretical, as well as
numerical and experimental scrutiny.

More recently, the RFIM has been realized in diluted dipolar magnets in
a transverse field such as ${\rm LiHo_xY_{1-x}F_4}$. However, most
experimental studies focus on diluted antiferromagnets in a field
(DAFF), such as ${\rm
Fe_{x}Zn_{1-x}F_2}$.\cite{belanger:98,barber:00,ye:02,ye:04,ye:06}
Fishman and Aharony\cite{fishman:79} were the first to note that a
random antiferromagnet in a field can be described by the RFIM, and
Cardy\cite{cardy:84a} predicted, using a mean-field argument, that the
critical behavior of both models should be in the same universality
class in the limit of small fields.  The work of Fishman and
Aharony,\cite{fishman:79} as well as Cardy,\cite{cardy:84a} therefore
opened the door for intense experimental investigation of the RFIM via
DAFF materials. However, early experiments and simulations already
hinted towards discrepancies between experimental and numerical
estimates of the critical
exponents.\cite{sourlas:99,hartmann:99c,belanger:98} On the other hand, exact
ground-state calculations using moderate system sizes suggested an
agreement between the critical exponents for both models when the random
fields are Gaussian distributed, however not when the random fields are
drawn from a bimodal distribution.\cite{sourlas:99,hartmann:99c}
This result, however, has been revised recently,\cite{fytas:13} i.e.,
the universality class of the RFIM is independent of the form of the 
implemented random-field distribution.

In this paper we perform detailed Monte Carlo simulations of both the
RFIM and the DAFF. The latter is studied at 30\% dilution, i.e., below
the percolation threshold for vacancies. Using a finite-size scaling
analysis of the Binder cumulant, the two-point finite-size correlation
function, and the susceptibility, we show that even in the limit of small
fields---where the Cardy mapping\cite{cardy:84a} is expected to
work---both models seem to be in different universality classes.
Therefore, care should be taken when making predictions for the critical
behavior of the RFIM using experiments on DAFF materials. Finally, we
present heuristic analytical expressions based on our numerical data for
the phase boundaries of both models to help guide experimental studies.

The manuscript is structured as follows. In Sec.~\ref{sec:models} we
introduce both the RFIM and the DAFF, followed by an explanation of
the used algorithms in Sec.~\ref{sec:alg}, as well as the measured
quantities in Sec.~\ref{sec:meth}. In Sec.~\ref{sec:num_res}
we show our numerical results, followed by a detailed discussion
of the phase boundaries and universality between both models in
Sec.~\ref{sec:discussion}.

\section{\label{sec:models}Models}

The Hamiltonian of the diluted antiferromagnet in a field (DAFF) is given by
\begin{equation} 
\mathcal{H}_\text{\tiny DAFF}
= +J\sum_{\langle i,j\rangle}
\varepsilon_i\varepsilon_j S_i S_j - 
B\sum_i \varepsilon_i S_i \; , 
\label{eq:ham_DAFF}
\end{equation}
and the Hamiltonian for the random-field Ising model (RFIM) is
\begin{equation}
\mathcal{H}_\text{\tiny RFIM} 
= -J \sum_{\langle i,j\rangle} S_i S_j 
- h\sum_i \delta_i S_i \; .
\label{eq:ham_RFIM}
\end{equation}
In Eqs.~\eqref{eq:ham_DAFF} and \eqref{eq:ham_RFIM} $S_i\in\{\pm 1\}$
represent Ising spins, $J = 1$ is the coupling constant between two
adjacent spins, and $\langle i,j\rangle$ denotes a sum over nearest
neighbors. The linear term in $S_i$ couples to an external field: For
the DAFF it is an externally-applied uniform field $B$, whereas for the
RFIM the spins couple to a random field of strength $h\delta_i$, where
the $\delta_i$ are quenched random variables chosen from a Gaussian
distribution with zero mean and standard deviation unity.  This means
that the typical field has strength $h$.  In the DAFF $\varepsilon_i \in
\{0,1\}$ represents the site dilution, where each site is randomly and
independently occupied by a spin ($\varepsilon_i=1$) with probability
$p$. Here, we fix the dilution to $1 - p = 0.3$.  Both models are
studied in three space dimensions on a lattice with $N = L^3$ spins, $L$
being the linear size of the lattice.

\section{\label{sec:alg}Algorithm}

The simulations are done using the Machta-Newman-Chayes replica-exchange
(MNC) algorithm\cite{machta:00} combined with single-spin Metropolis
Monte Carlo.\cite{newman:99,katzgraber:09e} The MNC algorithm is a
mixture of the Swendsen-Wang exchange algorithm\cite{swendsen:86} and
simulated tempering Monte Carlo.\cite{geyer:91,marinari:92} Note that
the latter is not efficient when simulating random-field
systems.\cite{moreno:03} The advantage of the MNC algorithm over
standard parallel tempering lies in the fact that we can choose any path
in the field--temperature plane.  Although parallel tempering can also
be implemented with a variable field, the method does not perform
efficiently when systems have disorder.\cite{young:04}

In the MNC algorithm\cite{machta:00} a cluster of connected spins is
grown between two replicas with the same disorder but at different
points in the parameter space, i.e., $(T,B)$ and $(T',B')$, where $T$
represents the temperature and $B$ the external field (here for the case
of the DAFF).  Starting from an arbitrary spin with different sign in
both realizations, adjacent spins pointing in the same direction are
successively added to the cluster with probability
\begin{equation}
p(\beta,\beta') = 1 - \exp\{-2(\beta+\beta')\} \; , 
\end{equation}
where $\beta = 1/T$ is the inverse temperature.  Once no more spins can
be added to the cluster $\mathcal{C}$, it flips with
the Metropolis probability\cite{comment:mistake} 
$\min\{1,\exp(-\Sigma)\}$, where
\begin{equation} 
\Sigma_\text{\tiny DAFF} = 2\text{sign}(\mathcal{C}) 
	\Big[(\beta - \beta')(n_{++} - n_{--})
		+ (B - B')|\mathcal{C}| \Big]
\end{equation}
for the DAFF, and for the RFIM
\begin{equation} 
\Sigma_\text{\tiny RFIM} = 2\text{sign}(\mathcal{C}) 
        \Big[(\beta - \beta')(n_{++} - n_{--})
                - (h - h') \sum_{i \in \mathcal{C}} \delta_i \Big].
\end{equation}
Here $|\mathcal{C}|$ is the number of spins in the cluster,
$\text{sign}(\mathcal{C})$ the orientation of the spin in the replica
having inverse temperature $\beta$, $n_{++}$ and $n_{--}$ are the number
of bonds connecting to nearest neighbors of the cluster with spin up and
spin down in both replicas, respectively. After each cluster update,
$(L/2)^3$ attempts to flip single spins are performed, where $L$ is the
linear size of the system.

As stated before, the MNC algorithm enables us to perform simulations
along any arbitrary path in parameter space. We denote such path a replica
chain (RC). The phase boundaries for the RFIM and DAFF in the
field--temperature plane are well described by ellipses (see below). To
reduce corrections to finite-size scaling\cite{wilding:92,joerg:08a} we
therefore choose paths in the field--temperature plane that cut the
phase boundaries at as orthogonal an angle as possible.  This means
that, in general, $T\sim h$ for the RFIM and $T\sim B$ for the DAFF. To
ensure efficient mixing and therefore fast convergence of the Monte
Carlo method, we additionally connect the point with the highest field
within the disordered phase to another RC that runs parallel to the
approximated phase boundary to a temperature $T > T_c$ and $B=0$ ($h=0$
for the RFIM), where $T_c$ is the critical temperature of the model at
zero field (see Fig.~\ref{fig:PB}, light dashed lines). This end point
of the second RC is simulated efficiently by the Wolff cluster
algorithm.\cite{wolff:89} Simulation parameters are listed in Tables
\ref{tab:rfim_samples} and \ref{tab:daff_samples} for the RFIM and DAFF,
for the first RCs, respectively.

\begin{table}[!ht]
\caption{
Simulation parameters for the RFIM along different nontrivial paths of
the type $h = a + bT$ in the $h$--$T$ plane for different linear system
sizes $L$ (the first two path types have $b=0$).  $N_{\rm sa}$ is the
number of disorder realizations.  $N_T$ corresponds to the number of
temperatures (points) along the simulation path. $T_{\rm min}$ and
$T_{\rm max}$ are the lowest and highest temperature simulated,
respectively. The equilibration/measurement times are $2^x$ Monte Carlo
sweeps.}
\label{tab:rfim_samples}
\begin{tabular*}{\columnwidth}{@{\extracolsep{\fill}} l r r c c c c}
\hline
\hline
simulation path &       $L$     & $N_{\rm sa}$ & $N_T$ &$T_{\rm min}$ &$T_{\rm max}$& $x$ \\\hline
$h=0.225$&        $8$     &$1536$ & $25$  &$4.00$& $5.00$&$18$\\
$h=0.225$&        $10$    &$827$  & $25$  &$4.00$& $5.00$&$18$\\
$h=0.225$&        $12$    &$2048$ & $17$  &$4.30$& $4.80$&$18$\\
$h=0.225$&        $16$    &$1024$ & $19$  &$4.35$& $4.70$&$18$\\
$h=0.225$&        $20$    &$1024$ & $19$  &$4.35$& $4.70$&$18$\\
$h=0.225$&        $24$    &$1024$ & $26$  &$4.40$& $4.69$&$18$\\
$h=0.225$&        $28$    &$666$  & $26$  &$4.40$& $4.69$&$18$\\
$h=0.225$&        $32$    &$406$  & $26$  &$4.40$& $4.69$&$18$\\
$h=0.225$&        $36$    &$1017$ & $26$  &$4.40$& $4.69$&$18$\\[1ex]
$h=0.5$&          $10$    &$2503$ & $17$  &$4.20$& $4.60$&$18$\\
$h=0.5$&          $12$    &$4035$ & $17$  &$4.20$& $4.60$&$18$\\
$h=0.5$&          $16$    &$2048$ & $17$  &$4.20$& $4.60$&$18$\\
$h=0.5$&          $20$    &$1024$ & $14$  &$4.30$& $4.50$&$18$\\
$h=0.5$&          $24$    &$512$  & $14$  &$4.30$& $4.50$&$18$\\[1ex]
$h=1.22T - 3.43$ &$10$    &$4096$ & $15$  &$3.40$& $4.10$&$18$\\
$h=1.22T - 3.43$ &$12$    &$3852$ & $15$  &$3.40$& $4.10$&$18$\\
$h=1.22T - 3.43$ &$16$    &$1177$ & $17$  &$3.65$& $4.10$&$18$\\
$h=1.22T - 3.43$ &$18$    &$862$  & $17$  &$3.65$& $4.10$&$18$\\
$h=1.22T - 3.43$ &$20$    &$957$  & $17$  &$3.60$& $4.00$&$18$\\
$h=1.22T - 3.43$ &$24$    &$976$  & $17$  &$3.60$& $4.00$&$18$\\
$h=1.22T - 3.43$ &$28$    &$646$  & $17$  &$3.60$& $4.00$&$18$\\
$h=1.22T - 3.43$ &$32$    &$379$  & $17$  &$3.60$& $4.00$&$18$\\[1ex]
$h=2.67T - 6.10$ &$8$     &$4071$ & $25$  &$2.80$& $3.06$ &$18$\\
$h=2.67T - 6.10$ &$10$    &$4045$ & $25$ &$2.80$& $3.06$&$18$\\
$h=2.67T - 6.10$ &$12$    &$512$  & $27$  &$2.85$& $3.00$&$18$\\
$h=2.67T - 6.10$ &$14$    &$512$  & $27$  &$2.85$& $3.00$&$18$\\
$h=2.67T - 6.10$ &$16$    &$605$  & $17$  &$2.85$& $2.95$&$18$\\
$h=2.67T - 6.10$ &$18$    &$1024$ & $27$ &$2.85$& $3.05$&$18$\\
$h=2.67T - 6.10$ &$20$    &$512$  & $31$  &$2.86$& $2.93$&$18$\\
$h=2.67T - 6.10$ &$22$    &$981$  & $31$  &$2.85$& $3.05$&$18$\\
$h=2.67T - 6.10$ &$24$    &$1024$ & $31$ &$2.85$& $3.05$&$18$\\[1ex]
$h=4.94T - 6.80$ &$16$    &$1912$ & $15$ &$1.76$& $1.88$&$18$\\
$h=4.94T - 6.80$ &$18$    &$2048$ & $15$ &$1.76$& $1.88$&$18$\\
$h=4.94T - 6.80$ &$20$    &$1858$ & $15$ &$1.76$& $1.89$&$18$\\
$h=4.94T - 6.80$ &$24$    &$906$  & $15$  &$1.76$& $1.89$&$18$\\
$h=4.94T - 6.80$ &$28$    &$505$  & $15$  &$1.76$& $1.89$&$18$\\
$h=4.94T - 6.80$ &$32$    &$627$  & $15$  &$1.76$& $1.89$&$18$\\
\hline
\hline
\end{tabular*}
\end{table}

\begin{table}[!ht]
\caption{
Simulation parameters for the DAFF along nontrivial paths of the type $B
= a + bT$ in the $B$--$T$ plane for different linear system sizes $L$
(the first two path types have $b=0$).  $N_{\rm sa}$ is the number of
disorder realizations. $N_T$ corresponds to the number of temperatures
(points) along the simulation path.  $T_{\rm min}$ and $T_{\rm max}$ are
the smallest and the highest temperatures of the RC, respectively. The
equilibration/measurement times are $2^x$ Monte Carlo sweeps.}
\label{tab:daff_samples}
\begin{tabular*}{\columnwidth}{@{\extracolsep{\fill}} l r r c c c c}
\hline
\hline
simulation path &       $L$     & $N_{\rm sa}$ & $N_T$ &$T_{\rm min}$ &$T_{\rm max}$ & $x$ \\\hline
$B=0.1$ & $8$   &$2166$ &$26$   &$2.50$& $3.50$  &$18$\\
$B=0.1$ & $12$  &$1208$ &$26$   &$2.50$& $3.50$&$18$\\
$B=0.1$ & $14$  &$1042$ &$18$   &$2.70$& $3.30$&$18$\\
$B=0.1$ & $16$  &$2048$ &$19$   &$2.80$& $3.30$&$18$\\
$B=0.1$ & $18$  &$1104$ &$19$   &$2.80$& $3.30$&$18$\\
$B=0.1$ & $20$  &$796$  &$21$    &$2.80$& $3.35$&$18$\\
$B=0.1$ & $24$  &$444$  &$21$   &$2.80$& $3.35$&$18$\\ 
$B=0.1$ & $28$  &$505$  &$21$   &$2.80$& $3.35$&$18$\\ 
$B=0.1$ & $32$  &$322$  &$21$   &$2.80$& $3.35$&$18$\\[1ex]
$B=1.0$ & $14$  &$1271$ &$21$   &$2.70$& $3.20$&$18$\\ 
$B=1.0$ & $16$  &$1718$ &$21$   &$2.70$& $3.20$&$18$\\ 
$B=1.0$ & $18$  &$1215$ &$21$   &$2.70$& $3.20$&$18$\\ 
$B=1.0$ & $20$  &$888$  &$21$   &$2.70$& $3.20$&$18$\\ 
$B=1.0$ & $24$  &$491$  &$21$   &$2.70$& $3.20$&$18$\\ 
$B=1.0$ & $28$  &$556$  &$21$   &$2.70$& $3.20$&$18$\\ 
$B=1.0$ & $32$  &$352$  &$21$   &$2.70$& $3.20$&$18$\\[1ex]
$B=0.2T$& $ 8$  &$1344$ &$17$   &$2.55$& $3.30$  &$18$\\ 
$B=0.2T$& $10$  &$685$  &$17$   &$2.55$& $3.30$&$18$\\ 
$B=0.2T$& $12$  &$452$  &$17$   &$2.55$& $3.30$&$18$\\ 
$B=0.2T$& $16$  &$542$  &$31$   &$2.87$& $3.50$&$18$\\ 
$B=0.2T$& $20$  &$1564$ &$31$  &$2.87$& $3.50$ &$18$\\ 
$B=0.2T$& $22$  &$825$  &$31$   &$2.87$& $3.50$&$18$\\ 
$B=0.2T$& $24$  &$189$  &$31$   &$2.87$& $3.50$&$18$\\ 
$B=0.2T$& $26$  &$128$  &$31$   &$2.87$& $3.50$&$18$\\ 
$B=0.2T$& $28$  &$115$  &$31$   &$2.87$& $3.50$&$18$\\ 
$B=0.2T$& $30$  &$558$  &$31$   &$2.87$& $3.50$&$18$\\ 
$B=0.2T$& $32$  &$383$  &$31$   &$2.87$& $3.50$&$18$\\[1ex]
$B=0.67T$& $10$  &$1201$ &$30$ &$2.45$& $3.50$ &$18$\\ 
$B=0.67T$& $12$  &$711$  &$30$  &$2.45$& $3.50$&$18$\\ 
$B=0.67T$& $16$  &$305$  &$30$  &$2.45$& $3.50$ &$18$\\ 
$B=0.67T$& $20$  &$512$  &$27$  &$2.35$& $3.50$ &$18$\\ 
$B=0.67T$& $22$  &$1024$ &$27$ &$2.35$& $3.50$&$18$\\ 
$B=0.67T$& $24$  &$2048$ &$30$ &$2.35$& $3.50$&$18$\\ 
$B=0.67T$& $28$  &$1024$ &$27$ &$2.35$& $3.50$ &$18$\\ 
$B=0.67T$& $32$  &$741$  &$30$  &$2.37$& $3.50$&$18$\\[1ex]
$B=1.5T$& $10$  &$1920$ &$17$   &$1.30$& $1.62$&$18$\\ 
$B=1.5T$& $12$  &$1984$ &$17$   &$1.30$& $1.62$&$18$\\ 
$B=1.5T$& $16$  &$2048$ &$17$   &$1.30$& $1.62$&$18$\\ 
$B=1.5T$& $18$  &$2048$ &$26$   &$1.30$& $3.50$&$18$\\ 
$B=1.5T$& $20$  &$1056$ &$20$   &$1.35$& $1.60$&$18$\\ 
$B=1.5T$& $24$  &$807$  &$20$   &$1.35$& $1.60$&$18$\\ 
$B=1.5T$& $28$  &$457$  &$20$   &$1.35$& $1.60$&$18$\\ 
$B=1.5T$& $32$  &$532$  &$20$   &$1.35$& $1.60$&$18$\\ 
$B=1.5T$& $36$  &$336$  &$20$   &$1.35$& $1.60$&$18$\\[1ex]
\hline
\hline
\end{tabular*}
\end{table}

Finally, we also study the DAFF at zero temperature using the method
introduced in Refs.~\onlinecite{esser:97} and \onlinecite{hartmann:98}.
Here, the DAFF is mapped onto a graph\cite{picard:75} with $N$ nodes
($N$ is the number of spins) attached to a source and a sink node, all
connected in a distinct manner via edges with positive edge weights. The
edge weights are calculated depending on the local staggered field,
i.e., $\pm B$. The maximum flow/minimum cut is obtained using the
algorithm introduced in Ref.~\onlinecite{goldberg:88}.  The minimum cut
is a direct representation of the ground-state spin configuration from
which derived quantities, such as a zero-temperature Binder ratio, can
be calculated.  Note that the method takes the ground-state degeneracy
into account. The simulation parameters for the DAFF at zero temperature
are shown in Table \ref{tab:daffT0_samples}.

\begin{table}[!ht]
\caption{
Simulation parameters for the DAFF at zero temperature for different
fields $B$ and for different linear system sizes $L$.  $N_{\rm sa}$ is
the number of disorder realizations.  $B_{\rm min}$ and $B_{\rm max}$
are the lowest and highest fields simulated, and $N_B$ corresponds to
the number of fields simulated to perform a finite-size scaling
analysis.
\label{tab:daffT0_samples}}
\begin{tabular*}{\columnwidth}{@{\extracolsep{\fill}} l r c c c}
\hline
\hline
$L$     & $N_{\rm sa}$ & $B_{\rm min}$ & $B_{\rm max}$ & $N_B$ \\\hline
$24$    & $10302$ & $2.00$& $ 4.30$& $31$  \\
$32$    & $2091$ 	& $2.40$& $ 2.70$& $16$  \\
$48$    & $2091$ 	& $2.10$& $ 2.80$& $17$  \\
$64$    & $2091$ 	& $2.30$& $ 2.70$& $21$  \\
$72$    & $2040$ 	& $2.30$& $ 2.54$& $17$  \\
$96$    & $5100$ 	& $2.30$& $ 2.54$& $17$  \\
$128$   & $3586$ & $2.30$& $ 2.47$& $22$  \\
\hline
\hline
\end{tabular*}
\end{table}

\section{\label{sec:meth}Observables}

Both the DAFF and RFIM undergo second-order phase transitions as a
function of temperature and field.  To pinpoint the transition
temperature, we measure the Binder cumulant,\cite{binder:81b} as well as
the two-point finite-size correlation
function.\cite{cooper:82,ballesteros:00,palassini:99b} To compute these
observables, we measure the magnetization per spin
\begin{equation} 
M=\frac{1}{N}\sum_i^N S_i \; .
\label{eq:m}
\end{equation}
For the DAFF we measure the {\em staggered} magnetization, i.e., each
second spin is counted opposite to its orientation in a
three-dimensional checker-board manner. For simplicity, we refer to the
staggered magnetization also as $M$. An antiferromagnetically-ordered
spin configuration has therefore $M=1$. A Binder cumulant for $M$ can 
then be defined via
\begin{equation} 
g(T,L) = \frac{1}{2} 
\left(3 - \frac{[\langle M^4 \rangle]_\text{\rm av}} 
{[\langle M^2 \rangle^2]_\text{\rm av}} \right)\; ,
\label{eq:Binder}
\end{equation}
where $\langle \cdots \rangle$ represents a thermal average and
$[\cdots]_\text{\rm av}$ an average over disorder (field or dilution
configurations) for a fixed value of $h$ (RFIM) or $B$ (DAFF).  Close to
criticality the Binder ratio scales as
\begin{equation} 
g(T,L) = {\tilde G}[L^{1/\nu}(T-T_c)] \; ,
\label{eq:binderRescale}
\end{equation}
where ${\tilde G}$ is a universal function.
Note that for the DAFF, when $T = 0$, $g(B,L) = {\tilde
G}^\prime[L^{1/\nu}(B-B_c)]$.  To compute the two-point finite-size
correlation function we first calculate the wave-vector-dependent
susceptibility (along the $x$ direction) via
\begin{equation}
\chi(\mathbf{k}) = 
\left[\left\langle 
    \left(
	\frac{1}{N}\sum_j S_j e^{i k x_j}
    \right)^2
\right\rangle\right]_\text{\rm av} \; .
\label{eq:susc}
\end{equation}
The two-point finite-size correlation function is then given by
\begin{equation}
\xi_L =	
\frac{1}{2\sin(k_\text{min}/2)} 
\sqrt{\frac{\chi(\mathbf{0})}{\chi(\mathbf{k_\text{min}})}-1}
\end{equation}
with ${\bf k}_\text{min} = (2\pi/L, 0, 0)$. The two-point finite-size
correlation function scales as
\begin{equation}
\xi_L(T,L)/L = {\tilde X} [L^{1/\nu}(T-T_c)] \; .
\label{eq:twoptcorrRescale}
\end{equation}
Using both the Binder ratio and the two-point finite-size correlation
function allows us to perform a detailed finite-size scaling analysis to
determine the critical exponent $\nu$, as well as to test if both models
share the same universality class using the method introduced in
Ref.~\onlinecite{katzgraber:06}.  To obtain an optimal data collapse we
use a Levenberg-Marquardt minimization combined with a bootstrap
analysis, see Ref.~\onlinecite{katzgraber:06}. This allows us to
determine the optimal values of the critical parameters $T_c$ and $\nu$
with a statistical error bar by fitting the data to a third-order
polynomial that approximates the scaling functions ${\tilde G}(x)$ and
${\tilde X}(x)$ close to $x = 0$, where $x=L^{1/\nu}(T - T_c)$.

Finally, to determine the critical exponent $\eta$, we determine the
peak position of the connected susceptibility given by
\begin{equation}
\chi = \frac{1}{T}
\left(
   \left[\langle M^2 \rangle\right]_{\rm av}
- 
   \left[\langle M \rangle\right]_{\rm av}^2
\right) \, ,
\label{eq:cc}
\end{equation}
where the magnetization $M$ is given by Eq.~\eqref{eq:m}.  Note that the
connected susceptibility is related to Eq.~\eqref{eq:susc} in the limit
of zero wave vector. Furthermore, in the thermodynamic limit $[\langle M
\rangle]_{\rm av} = 0$ for $T = T_c$ so, in principle,
Eq.~\eqref{eq:susc} could also be used for the analysis. In general, the
susceptibility scales as
\begin{equation}
\chi \sim L^{2 - \eta} \widetilde{C}
[L^{1/\nu}(T - T_c)] \, .
\label{eq:scalchi}
\end{equation}
Therefore, when $T = T_c$ the function $\widetilde{C}$ is a constant
independent of the system size and $\chi \sim  L^{2 - \eta}$ from which
the exponent $\eta$ can be determined.

\section{\label{sec:num_res} Results}

\begin{figure*}[htp]

\includegraphics[width=0.45\textwidth,angle=270]{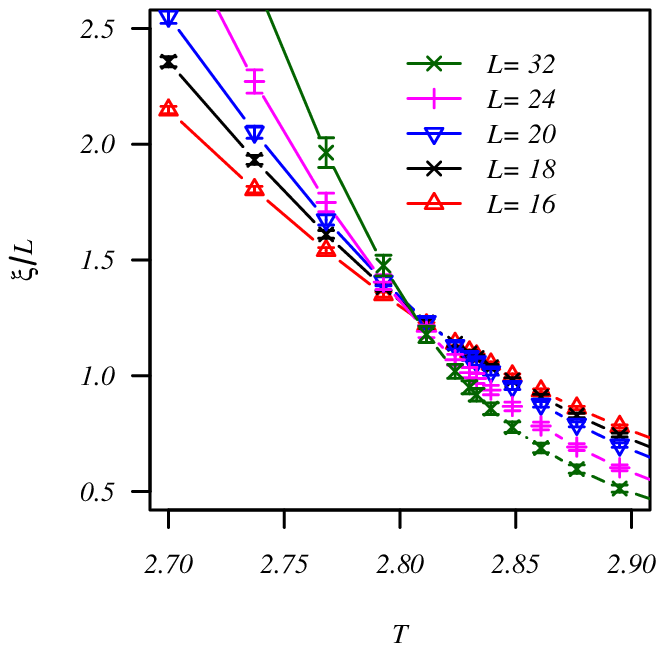}
\hspace*{2em}
\includegraphics[width=0.45\textwidth,angle=270]{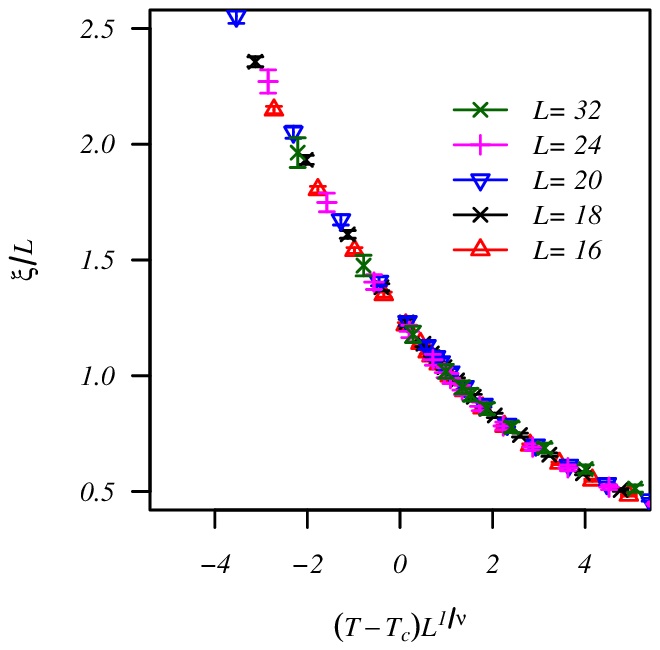}

\caption{(Color online)
Left: Two-point finite-size correlation function $\xi_L/L$ vs
temperature $T$ for the DAFF with $B = 1.0$ and different linear system
sizes $L$.  Finite-size corrections are small and the data cross at one
point signaling a transition.  Right: Finite-size scaling collapse of
the data in the left panel. The best collapse is obtained with $T_c
\approx 2.807$ and $\nu \approx 1.2$.
\label{fig:B10_xi_T}}
\end{figure*}

\begin{figure*}[htp]

\includegraphics[width=0.45\textwidth,angle=270]{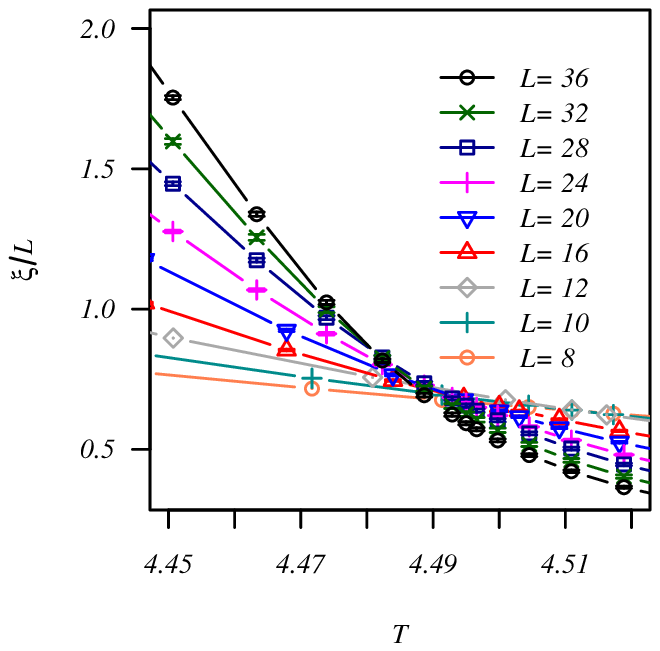}
\hspace*{2em}
\includegraphics[width=0.45\textwidth,angle=270]{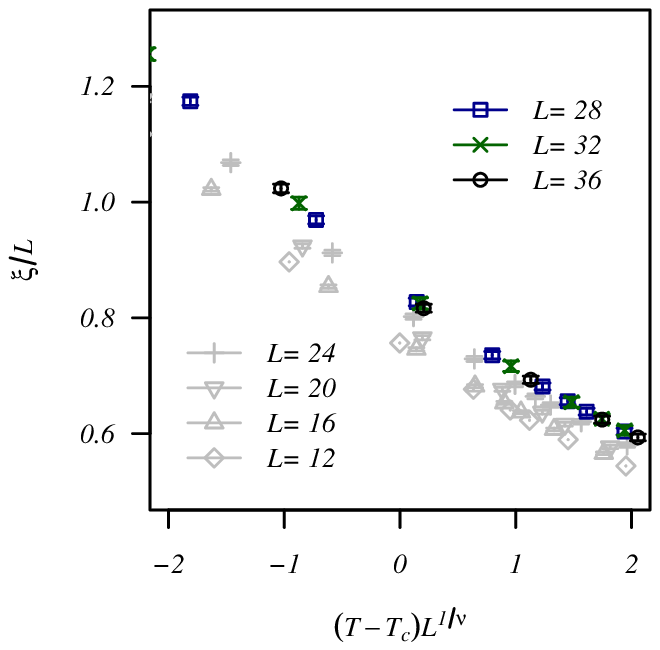}

\caption{(Color online)
Left: Two-point finite-size correlation function $\xi_L/L$ vs
temperature $T$ for the RFIM with $h = 0.225$ and different linear
system sizes $L$. Finite-size corrections are large.  Right: Finite-size
scaling collapse of the data in the left panel. Because of the large
corrections to scaling, only data for $L \ge 28$ are used in the scaling
collapse. Data for $L \le 24$ (light shaded) are not included in the data
collapse and shown to illustrate the corrections to scaling. The best
collapse is obtained with $T_c \approx 4.481$ and $\nu \approx 1.39$.
\label{fig:h0225_xi_T}}
\end{figure*}

The critical parameters for both the RFIM and the DAFF have been
computed via a finite-size scaling analysis of the two-point finite-size
correlation function [Eq.~(\ref{eq:twoptcorrRescale})] along the
different simulation paths. Finite-size corrections can be large for
small system sizes and are strongly field dependent, which is why for
some external fields in both models we do not include small systems in
the finite-size scaling analysis used to determine the critical
parameters. To illustrate the typical behavior, in
Fig.~\ref{fig:B10_xi_T}, left panel, we show the two-point finite-size
correlation function for the DAFF for $B = 1.0$ and different system
sizes. The data cross at a point, therefore signaling the existence of a
phase transition. Note that for this particular field corrections to
scaling are manageable and the data scale well, as can be seen in
Fig.~\ref{fig:B10_xi_T}, right panel. However, this is not always the
case, especially when the external field is large. For the RFIM
corrections to scaling are considerably stronger, even at small fields,
see Fig.~\ref{fig:h0225_xi_T}.

Using finite-size scaling we determine the location of the critical
points, as well as the associated critical exponent $\nu$ for the
different simulation paths. In addition, we also compute the critical
exponent $\eta$ by studying the finite-size behavior of the
susceptibility peak.  Data for the RFIM are summarized in Table
\ref{tab:results_rfim}, for the DAFF in Table \ref{tab:results_daff}.

\begin{table}[bph]
\caption{
Critical temperature $T_c$ and critical field $h_c$ computed from a
finite-size scaling analysis of the two-point finite-size correlation
function for the RFIM. $\nu$ is the critical exponent of the correlation
length. The exponent $\eta$ is computed from the peak of the susceptibility.
\label{tab:results_rfim}}
\begin{tabular*}{\columnwidth}{@{\extracolsep{\fill}}l l l l l}
\hline
\hline
simulation path    &  $T_c$	& $h_c$	    & $\nu$	  & $\eta$	\\
\hline
$h = 0.225$	   & $4.481(1)$	& $0.225$   & $1.39(4)$	  & $0.082(1)$\\ 
$h = 0.5$ 	   & $4.381(2)$ & $0.5$     & $1.30(5)$   & $0.202(16)$\\
$h = 1.22T - 3.4$  & $3.76(2)$	& $1.16(3)$ & $1.39(5)$   & $0.92(40)$\\
$h = 2.70T - 6.1$  & $2.89(5)$	& $1.7(1)$  & $1.3(1)$    & $0.47(15)$\\
$h = 4.94T - 6.8$  & $1.79(1)$	& $2.01(5)$ & $1.4(1)$    & $0.85(4)$\\
\hline
\hline
\end{tabular*}
\end{table}

\begin{table}[h] 
\caption{
Critical temperature $T_c$ and critical field $B_c$ computed from a
finite-size scaling analysis of the two-point finite-size correlation
function for the DAFF. $\nu$ is the critical exponent of the correlation
length.  The exponent $\eta$ is computed from the peak of the
susceptibility. Note that estimating $\eta$ was not possible for $B = 1.5$.
The last line lists data from zero-temperature simulations (see text). 
The estimate of the critical field $B_c$ is obtained from a finite-size 
scaling analysis of the zero-temperature Binder ratio.
\label{tab:results_daff}}
\begin{tabular*}{\columnwidth}{@{\extracolsep{\fill}}l l l ll }
\hline
\hline
simulation path & $T_c$		& $B_c$ 	& $\nu$		& $\eta$\\ 
\hline
$B = 0.1$	& $2.977(1)$	& $0.1$		& $1.34(5)$ 	& $0.406(26)$ \\
$B = 1.0$	& $2.807(1)$	& $1.0$		& $1.2(2)$ 	& $0.023(12)$ \\
$B = 0.2T$ 	& $2.908(4)$	& $0.582(8)$	& $1.36(7)$	& $0.11(2)$ \\
$B = 0.67T$ 	& $2.42(1)$ 	& $1.61(1)$ 	& $1.5(3)$	& $0.67(5)$ \\
$B = 1.5$ 	& $1.46(9)$ 	& $2.2(1)$ 	& $1.4(3)$ 	&  --- \\
$T = 0$		& $0$		& $2.32(2)$	& $1.43(2)$	& $0.68(1)$ \\
\hline
\hline
\end{tabular*}
\end{table}

\begin{figure}[htb]
\includegraphics[width=0.45\textwidth,angle=270]{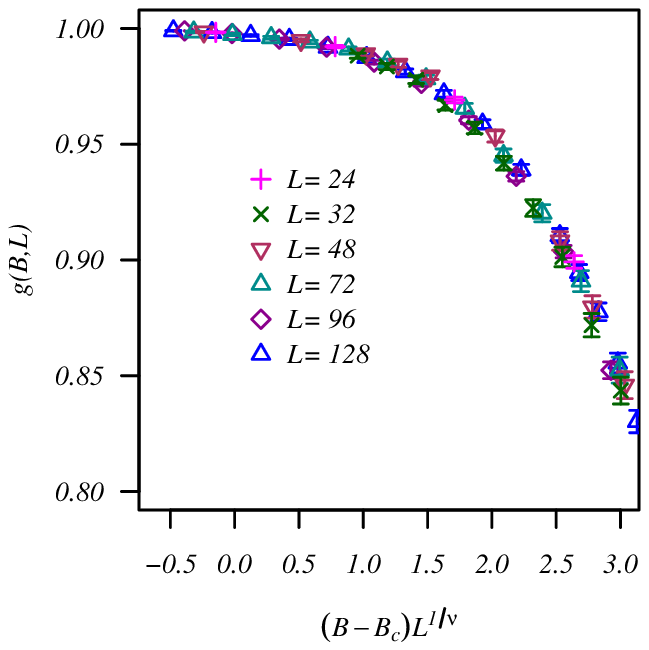}
\caption{(Color online) 
Data collapse of the zero-temperature Binder cumulant of the DAFF as a
function of the reduced scaling variable $(B - B_c)L^{1/\nu}$ for
different system sizes. The best collapse is obtained for $B_c^0\approx
2.32$ and $\nu \approx 1.43$.
\label{fig:GS_g}}
\end{figure}

To determine the critical field $B_c$ at zero temperature for the DAFF
we compute ground states with the algorithm introduced in
Ref.~\onlinecite{hartmann:98}. The same finite-size scaling technique as
used for the two-point finite-size correlation function (see above) can
be used to analyze the ground-state Binder cumulant. The data collapse
is shown in Fig.~\ref{fig:GS_g}. The results for the critical point and
the correlation-length exponent at zero temperature are stated in the
last line of Table \ref{tab:results_daff}.

\begin{figure*}[h]

\includegraphics[width=0.45\textwidth,angle=270]{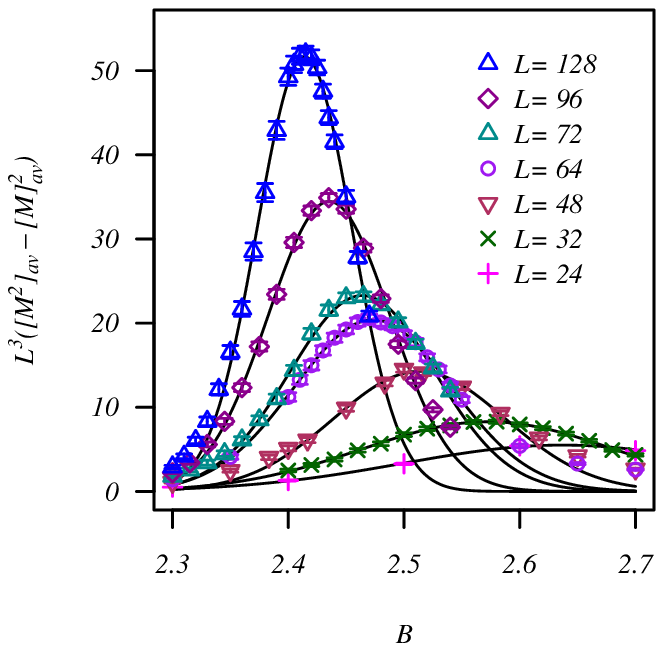}
\hspace*{2em}
\includegraphics[width=0.45\textwidth,angle=270]{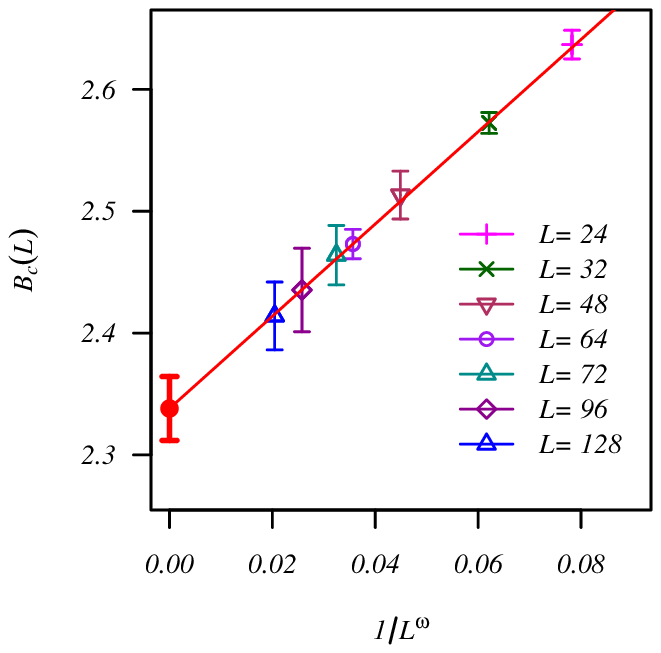}

\caption{(Color online)
Left: Fluctuations of the staggered magnetization of the DAFF as a
function of applied field $B$ for different system sizes. The peak
positions signals the presence of a transition. The data are well
described by a Gaussian close to the peak (solid lines). To determine the
thermodynamic critical field $B_c$ we extrapolate the data to infinite
system size (right panel) using $B_c(L) = B_c + a L^{-\omega}$. The best
fit is obtained for $B_c = 2.34(2)$ and $\omega = 1.25(9)$.  The red
(filled) point represents the thermodynamic extrapolation, $B_c = 2.34(2)$.
\label{fig:GS_F_B}}
\end{figure*}

\begin{figure*}[h]

\includegraphics[width=0.45\textwidth,angle=270]{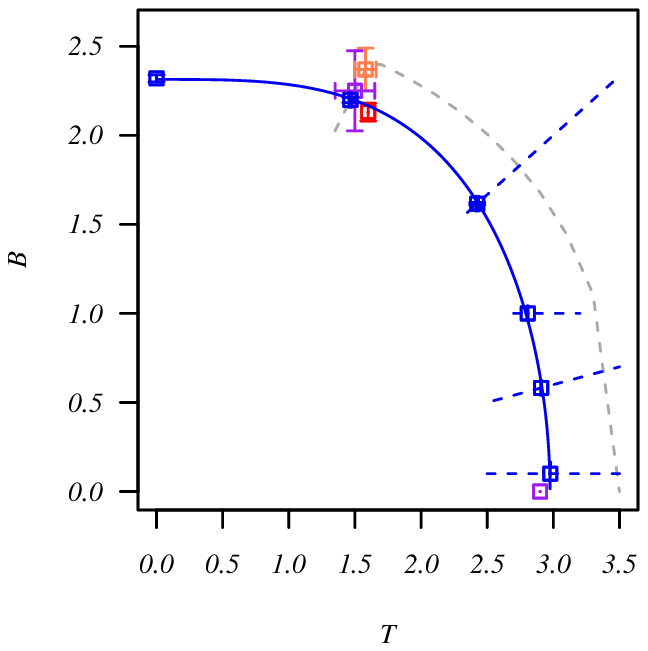}
\hspace*{2em}
\includegraphics[width=0.45\textwidth,angle=270]{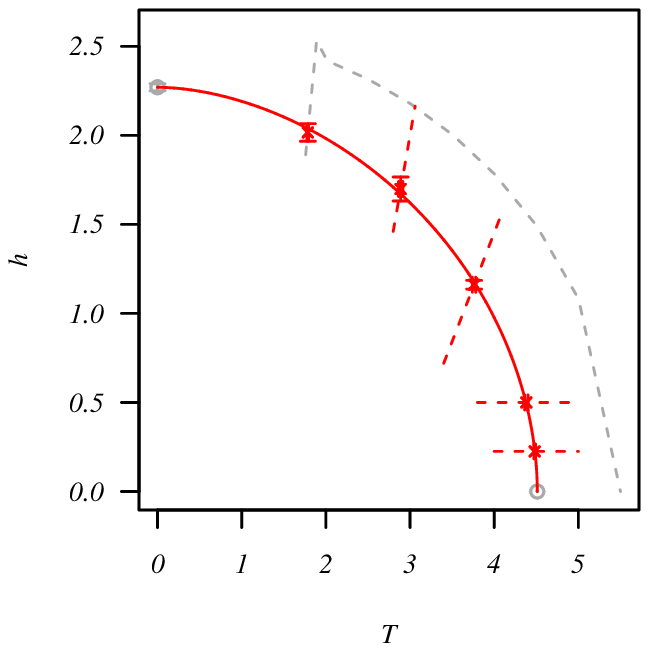}

\caption{(Color online)
Left: Empirical phase boundary of the DAFF ($p=0.7$). The red point is
from Ref.~\onlinecite{fernandez:11}, the coral point from
Ref.~\onlinecite{maiorano:07}, and the purple points from
Ref.~\onlinecite{ogielski:86}. Our data agree within error bars with
these previous studies. The blue (solid) curve is given by
Eq.~\eqref{eq:pb_daff}. The dashed lines represent the parts of the
simulation paths that cross the phase boundary. The light shaded line is
an example of the second RC that runs parallel to the approximated phase
boundary to a temperature $T > T_c$ and $B=0$ to speed up equilibration.
Right: Empirical phase boundary of the RFIM. The zero-field critical
temperature is $T_c^0=4.5115$\cite{talapov:96} and
$h_c^0=2.270$\cite{middleton:02} (gray open circles). The red (solid)
curve is given by Eq.~\eqref{eq:pb_rfim}. The dashed lines represent the
parts of the simulation paths that cross the phase boundary.  Again, the
light shaded line shows an example of the second RC that runs parallel
to the approximated phase boundary to a temperature $T > T_c$ and $h=0$
to speed up equilibration.
\label{fig:PB}}
\end{figure*}

We also determine the peak position of the fluctuations of the staggered
magnetization of the ground states:
\begin{equation}
\mathcal{F}(B) = 
	L^3\left([M^2]_{\rm av} - [M]_{\rm av}^2\right) \; .
\end{equation}
This approach has proven to be quite accurate in previous studies for
the susceptibility.\cite{ahrens:11} Because the fluctuations peak at the
putative transition, we fit a Gaussian to the peak and determine its
precise location. Error bars are determined via a configurational
bootstrap analysis.\cite{hartmann:09} Figure \ref{fig:GS_F_B}, left
panel, shows the fluctuations at zero temperature and as a function of
the applied field $B$. The peaks are well described by Gaussians. The
right panel of Fig.~\ref{fig:GS_F_B} shows an extrapolation of the peak
position to infinite system size assuming the functional form $B_c(L) =
B_c + a L^{-\omega}$. The best fit is obtained for $B_c = 2.34(2)$
[$\omega = 1.25(9)$], in agreement with the estimate using the Binder
cumulant, see Table \ref{tab:results_daff}.

Combining the data in Table \ref{tab:results_daff} with some values from
the literature\cite{ogielski:86,maiorano:07,fernandez:11,yllanes:11} we can
approximate to good accuracy the phase boundary for the DAFF via
\begin{equation}
\left(\frac{B_c}{\tilde B_c^0}\right)^{1.81} 
+ 
\;\;\;\;\;\;\;\;
\left(\frac{T_c}{\tilde T_c^0}\right)^{3.54}
= 1
\label{eq:pb_daff}
\end{equation}
with $\tilde T_c^0 \approx 2.980$  and $\tilde B_c^0 \approx 2.31$.
Similarly, using the data from Table \ref{tab:results_rfim} and known
values from the literature\cite{talapov:96,hartmann:01c,middleton:02} 
we obtain for
the RFIM
\begin{equation}
\left(\frac{h_c}{h_c^0}\right)^{1.95} 
+ 
\;\;\;\;\;\;\;\;
\left(\frac{T_c}{T_c^0}\right)^{1.80}
= 1
\label{eq:pb_rfim}
\end{equation}
with $h_c^0 = 2.27$\cite{middleton:02} and $T_c^0 =
4.5115$.\cite{talapov:96} Note that the critical phase boundary points
$T_c^0$ and $h_c^0$ have been determined to high precision in the
literature; see Refs.~\onlinecite{talapov:96} and
\onlinecite{middleton:02}, respectively. Furthermore, for the RFIM with
bimodal disorder, a similar elliptical phase diagram has been proposed in
Ref.~\onlinecite{fytas:08}. For the DAFF, $\tilde T_c^0$ and $\tilde
B_c^0$ are approximated but agree with the numerical estimates we
present.  In Fig.~\ref{fig:PB} we show the phase boundaries for the DAFF
(left panel) and the RFIM (right panel), together with the simulated
critical points. The dashed lines represent the simulation paths taken.

\begin{figure}[htb]
\includegraphics[width=0.5\textwidth,angle=270]{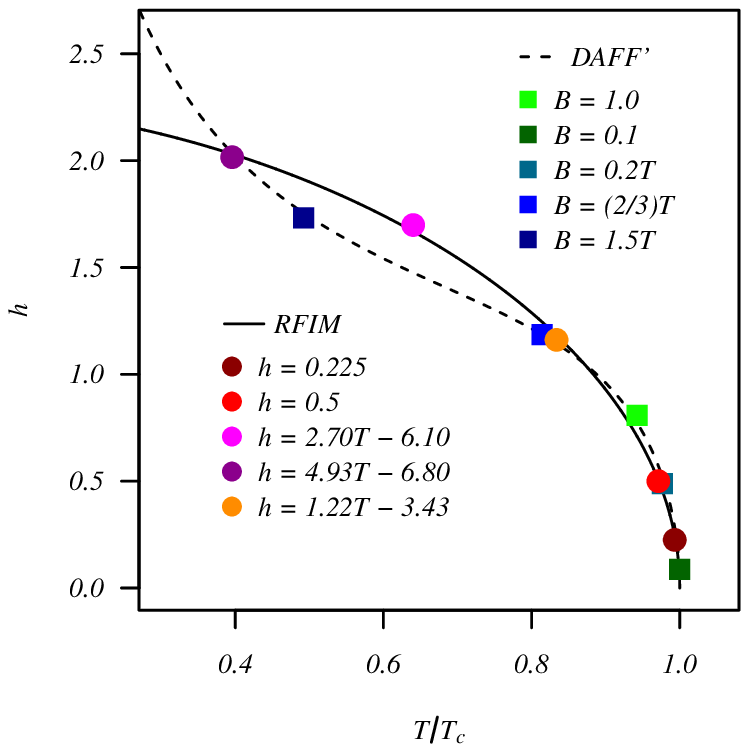}
\caption{(Color online)
Phase boundary of the RFIM (solid line, from Fig.~\ref{fig:PB}, right
panel). The temperature axis has been normalized with $T_c(h=0) =
4.5115$. The circles represent the different estimates of the critical
points along the different simulation paths.  The dashed line is the
phase boundary computed by applying Eq.~\eqref{eq:mapping} to the data
of the DAFF. Squares represent the different critical points simulated
for the DAFF along the different simulation paths.  An approximate
correspondence between the phase boundaries only works for fields $h
\lesssim 1.2$ ($B \lesssim 1.6$ for the DAFF).
\label{fig:cardy_mapping}}
\end{figure}

\begin{figure}[htb]
\includegraphics[width=0.5\textwidth,angle=270]{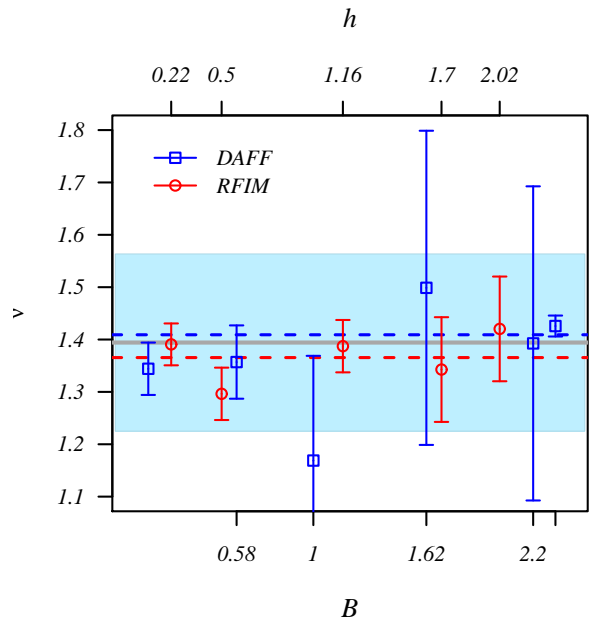}
\caption{(Color online)
Critical exponent $\nu$ as a function of the field $h$ (RFIM) and $B$
(DAFF).  The labels on the upper axis correspond to the random-field
strength $h$ (RFIM), those on the lower axis  to the external field $B$
(DAFF). The weighted mean is $\nu=1.39(17)$ (gray line) and the weighted
error is represented by the shaded (light blue) area. The difference
between $\bar \nu_\text{DAFF} = 1.41(15)$ (blue dashed line) and $\bar
\nu_\text{RFIM} = 1.37(12)$ (red dashed line) is marginal in comparison
to the error-bars of the data points. The RFIM ground-state value is
taken from Ref.~\onlinecite{middleton:02}.
\label{fig:nu_TTc}}
\end{figure}

\begin{figure}[htb]
\includegraphics[width=0.5\textwidth,angle=270]{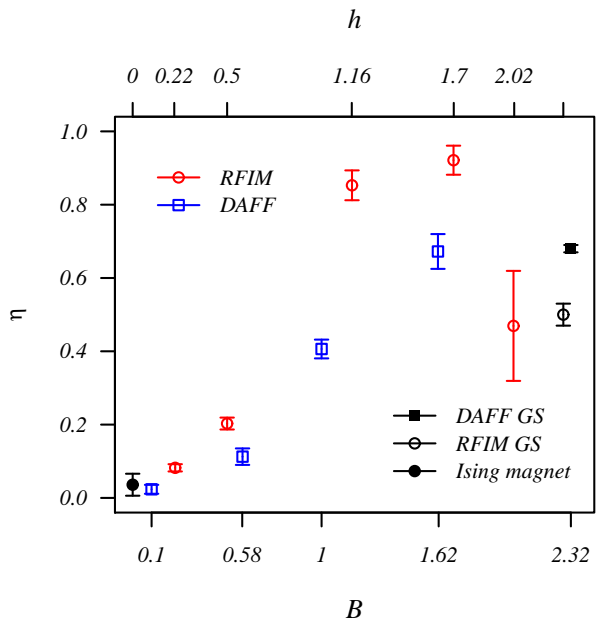}
\caption{(Color online)
Critical exponent $\eta$ as a function of the field $h$ (RFIM) and $B$
(DAFF).  The labels on the upper axis correspond to the random-field
strength $h$ (RFIM), those on the lower axis  to the external field $B$
(DAFF).  For comparison, we also add the estimates for the
three-dimensional Ising ferromagnet (filled circle at $h = 0$, marked
with `Ising magnet'),\cite{talapov:96} the RFIM at $T = 0$ (open circle
at $h = h_c$, marked with `RFIM GS'),\cite{hartmann:01c} and the DAFF at
$T = 0$ and $B = B_c$ computed from our ground-state data [$\eta(T=0)
\approx 0.68(1)$, filled square, marked with `DAFF GS']. Note that we
find very large fluctuations, i.e., a detailed determination of the
different universality classes is difficult.
\label{fig:eta}}
\end{figure}

\begin{figure*}[htb]
\includegraphics[height=0.96\textwidth,angle=270]{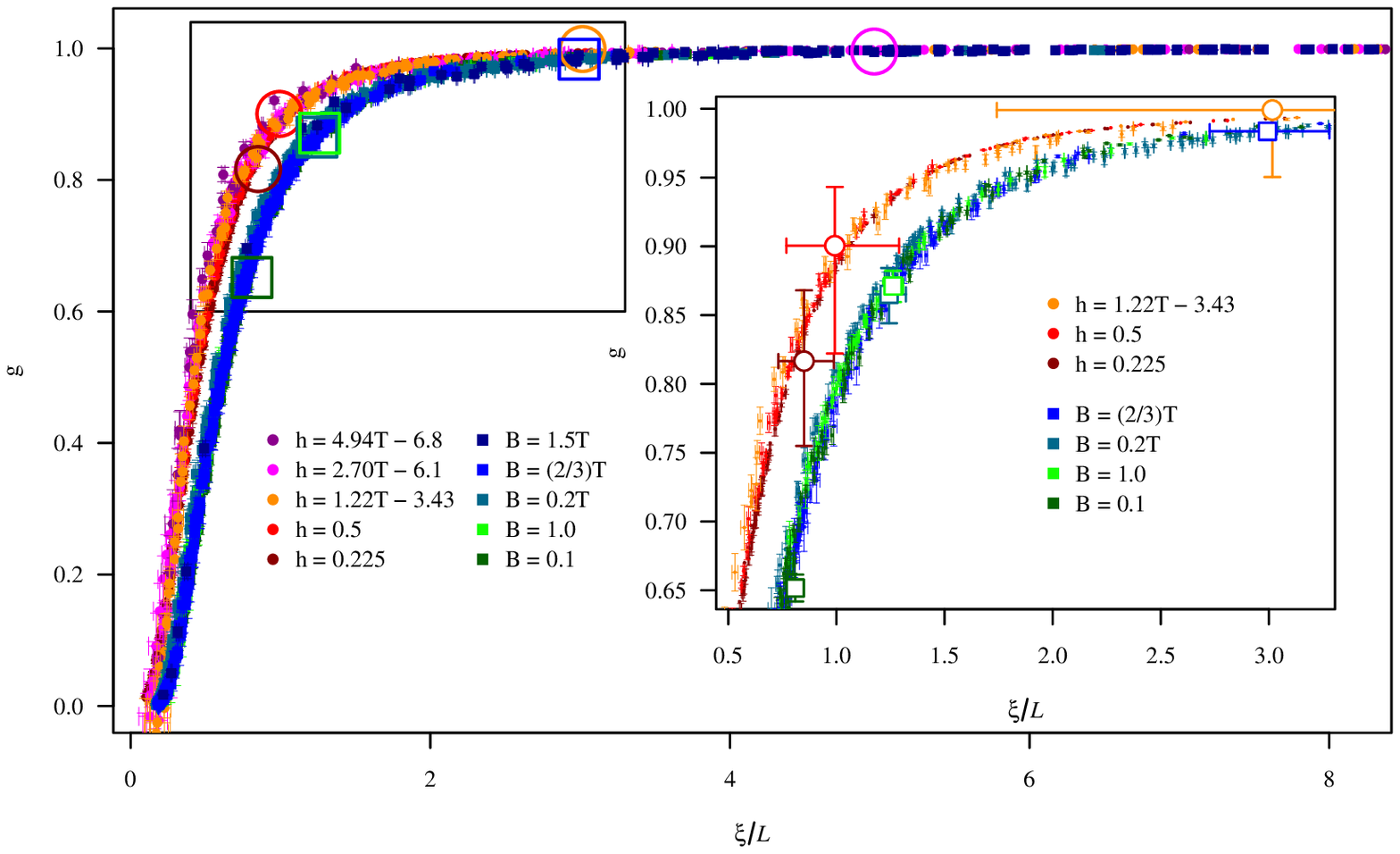}
\caption{(Color online)
Binder ratio $g$ as a function of the two-point finite-size correlation
function divided by the system size $\xi_L/L$ for several system sizes
and simulation paths.  Note that also small system sizes are included,
i.e., corrections to finite-size scaling are small.  The data for the
RFIM and DAFF collapse onto two distinct set of curves, suggesting that
both models do not share the same universality class.  The left set of
points (reddish/light tones, circles) are for the RFIM. The right set of
points (greenish/dark tones, squares) are for the DAFF. The large
circles for the RFIM (large squares for the DAFF) represent our
estimates of $g(\xi_L/L)$ at $T = T_c$.  The inset zooms into the
important region (large box), where the Cardy mapping should apply.
Clearly, both data sets are different, suggesting that the RFIM and the
DAFF do not share the same universality class.
\label{fig:universe}}
\end{figure*}

\section{\label{sec:discussion} Discussion}

Cardy\cite{cardy:84a} predicted an equivalence between the DAFF and the
RFIM for small applied fields using a mean-field argument.  This
equivalence is often quoted in experimental studies where materials
which are diluted antiferromagnets in a field are then described using
the RFIM (see, for example,
Refs.~\onlinecite{belanger:98,barber:00,ye:02,ye:04,ye:06}).

Equation (15) in Ref.~\onlinecite{cardy:84a} maps the RFIM onto the
DAFF:
\begin{equation}
h(T) = \frac
{p(1-p)(T_c^{\rm pure}/T)^2(B/T)^2 }
{( 1 - \theta^{MF}/T)^2}\;.
\label{eq:mapping}
\end{equation}
Here, $p=0.7$, $T_c^{\rm pure} = 4.5115$, and $\theta^{MF} = 2dJ = 6$ is
the mean-field coupling strength. We can now use the obtained phase
boundaries [Eqs.~\eqref{eq:pb_daff} and \eqref{eq:pb_rfim}] to compare
both models.  Figure \ref{fig:cardy_mapping} shows the phase boundary
for the RFIM (solid line, the circles represent the obtained critical
points along the different simulation paths) together with the phase
boundary for the DAFF mapped onto the RFIM space using
Eq.~\eqref{eq:mapping} (dashed line, the squares represent the obtained
critical points along the different simulation paths for the DAFF).  For
random-field strengths of up to $h\approx 1.2$---which means field
strengths of up to $B\approx 1.6$ for the DAFF---there is an approximate
correspondence between both models. However, as the figure clearly
illustrates, strictly speaking the correspondence only seems to work in
the limit of $h \to 0$ ($h \lesssim 0.3$).  Given the mean-field nature
of the Cardy argument, the agreement of the phase boundaries is rather
good. On the other hand, it is not surprising that for larger disorder,
they do not agree exactly.  It is of importance to take these
limitations of the Cardy mapping\cite{cardy:84a} into account when
studying diluted antiferromagnets in an external field experimentally
while attempting to describe the data analytically using the RFIM.
Furthermore, a basic finite-size scaling analysis leads to no systematic
deviations of the correlation-length exponent $\nu$. Including the
estimates for rough simulations at high fields, our results support
\begin{equation}
\nu=1.39(17) 
\end{equation}
for the range of fields studied, in agreement with previous studies,
such as $\nu_{\rm RFIM}=1.37(9)$, \cite{middleton:02}
$\nu=1.20(5)$\cite{ye:04} from experiments on ${\rm Fe}_{0.85}{\rm
Zn}_{\rm 0.15}{\rm F}_{2}$ ($p = 0.85$), or $\nu = 1.40(6)$ from the
disconnected part of the susceptibility of ${\rm Fe}_{0.93}{\rm Zn}_{\rm
0.07}{\rm F}_{2}$ ($p = 0.93$).\cite{slanic:01} Note that our results
are also compatible with the value $\nu = 1.10(15)$ computed by
Fernandez {\em et al.}~\onlinecite{fernandez:11} obtained for their
largest system size using the quotient method. They do find other
values of $\nu$ for smaller system sizes.  Our results are summarized in
Fig.~\ref{fig:nu_TTc}. As can be clearly seen, the difference between
the estimates for the critical exponent of the correlation length for
both models is marginal and within error bars: The {\em average}
estimate for the RFIM is $\bar \nu_\text{RFIM} = 1.37(12)$ (red line in
Fig.~\ref{fig:nu_TTc}), whereas for the DAFF $\bar \nu_\text{DAFF} =
1.41(15)$ (blue line in Fig.~\ref{fig:nu_TTc}).  This apparent agreement
of the critical exponent is quite good, given that the proposed
equivalence is based on a mean-field argument that typically leads to
quite different exponents compared to the true non-mean-field values.

However, the error bars are large and therefore a more detailed study
needs to be performed. To truly discern if both models are in the same
universality class, in addition to having one (apparently) agreeing
critical exponent, one would have to compute a second critical exponent.
We also analyzed the behavior of the magnetic susceptibility $\chi$
which has a peak at the phase transition.  By studying the finite-size
behavior of the peak height (not shown), we determine the critical
exponent $\eta$ using the finite-size scaling form of the
susceptibility, Eq.~\eqref{eq:scalchi}.  Our estimates of the critical
exponent $\eta$ along the phase boundary are shown in Fig.~\ref{fig:eta}
and summarized in Tables \ref{tab:results_rfim} and
\ref{tab:results_daff} for the RFIM and DAFF, respectively.
Fluctuations are very large, especially for large fields, but suggest
that both the RFIM and the DAFF might not share the same universality
class. For the DAFF, a clear systematic trend is visible that shows that
$\eta$ might be strongly field dependent for $B \gtrsim 1.6$, i.e., in
the curved portion of the phase boundary. However, note that the
exponent $\eta$ is very difficult to compute, as recently shown in
Ref.~\onlinecite{fytas:13}.  A different approach is the computation of
the critical exponent $\alpha$ that describes the divergence of the
specific heat.  However, for both the RFIM and the DAFF $\alpha$ is
close to zero.\cite{hartmann:98,middleton:02} Therefore, simulations of
very large system sizes that are currently not accessible numerically
are required.

Fortunately, there is a simple yet more sensitive method to verify if
two different systems share the same universality class without having
to compute any critical exponents:\cite{joerg:06,katzgraber:06} Both the
Binder cumulant and the two-point finite-size correlation function
divided by the system size are dimensionless quantities. By plotting one
as a function of the other, nonuniversal quantities cancel
out.\cite{katzgraber:06} For a given system, once large enough system
sizes are reached such that corrections to scaling are negligible, the
data for all system sizes collapse onto a universal curve within error
bars.  If two systems share the same critical exponent $\nu$, we expect
that all data should collapse onto the same universal curve within error
bars and, in particular, that the estimates of the Binder cumulant and
the two-point finite-size correlation function agree at the putative
critical point(s). We therefore would expect that data sets of
$g(\xi_L/L)$ for both the DAFF and the RFIM should agree for all
simulated temperatures and, in particular, for $T = T_c$.

Figure \ref{fig:universe} shows the Binder cumulant as a function of the
two-point finite-size correlation function divided by the system size
for both the DAFF and the RFIM. The left set of points (reddish/light
tones, circles) are for the RFIM. Data for the different simulation
paths used collapse onto a master curve. The right set of points
(greenish/dark tones, squares) are for the DAFF. Again, all data
collapse onto a master curve for all simulation paths taken.  This shows
that for this type of analysis the finite-size corrections are small for
both models and within the statistical fluctuations.  However, the data
sets for the RFIM and the DAFF do not agree, except in the trivial limit
where $g(T) \to 1$.  The large circles for the RFIM (squares for the
DAFF) represent our estimates of $g(\xi_L/L)$ at $T = T_c$. As can be
seen, the data for both models do not agree (i.e., a large circle should
sit on top of a large square), something which is even more clear when
zooming into the boxed area (inset). Note that the large error bars are
due to the uncertainty of the critical temperature.  This discrepancy
reveals the differences between the DAFF and the RFIM which could not be
detected within the scope of a mean-field calculation.

\section{\label{sec:summ}Conclusions}

We have performed extensive Monte Carlo simulations of the diluted
antiferromagnet in a field at 30\% dilution ($p=0.7$) and the
random-field Ising model. Using these data we show that the phase
boundaries for both models are well described by ellipses (see
Fig.~\ref{fig:PB}). In addition, using zero-temperature heuristic
methods, we compute the zero-temperature critical point for the DAFF
with 30\% dilution ($p=0.7$). We expect that the phase boundary for
other dilutions will be similar, albeit with different nonuniversal
parameters.

Furthermore, we numerically study the equivalence of the RFIM and the
DAFF as predicted by Cardy.\cite{cardy:84a} Our results show that only in
the limit of small fields do both phase boundaries map onto each other.

Finally, we perform a finite-size scaling analysis to determine the
critical exponent $\nu$ of the correlation length. Our results from the
two-point finite-size correlation function suggest that the exponent
$\nu$ agrees within error bars for both the RFIM and the DAFF. However,
error bars are large. To circumvent this problem, we study the Binder
cumulant as a function of the two-point finite-size correlation function
divided by the system size and show that both models apparently do not
share the same universality class. A computation of the exponent $\eta$
is extremely difficult and plagued by finite-size effects. Clearly, more
detailed simulations need to be performed to fully discern the critical
behavior of both models and fully determine their universality classes.
It would be interesting to also measure the critical behavior of the
specific heat (critical exponent $\alpha$).  However, because the
exponent is close to zero for both models, large system sizes are
needed; sizes that are currently not accessible via simulations.  We
conclude by cautioning researchers when using the equivalence of both
models.

\begin{acknowledgments} 

We would like to thank D.~P.~Belanger, M.~Niemann, and A.~P.~Young for
the fruitful discussions. H.G.K.~acknowledges support from the SNF
(Grant No.~PP002-114713) and the NSF (Grant No.~DMR-1151387).  We would
like to thank ETH Zurich for CPU time on the Brutus cluster, Texas A\&M
University for CPU time on the Eos cluster, as well as the
C.~v.~O.~Universit\"at Oldenburg for CPU time on the Hero cluster funded
by the DFG (INST 184/108-1 FUGG) and the ministry of Science and Culture
(MWK) of the Lower Saxony State.

\end{acknowledgments} 

\bibliography{refs,comments}

\begin{thebibliography}{53}
\bibitem[{com()}]{comment:mistake}
\bibinfo{note}{Note that Ref.~\onlinecite{machta:00} contains a mistake
  regarding the definition of $n_{++}$ and $n_{--}$.}
\end{thebibliography}

\end{document}